\documentclass[epj]{svjour}
\usepackage{graphics}
\usepackage[parfill]{parskip}    % Activate to begin paragraphs with an empty line rather than an indent
\usepackage{authblk}
\usepackage{caption}
\usepackage{subcaption}
\usepackage{graphicx}
\usepackage{float}
\usepackage[numbers]{natbib}

\usepackage{amssymb}
\usepackage{tensor}
 \usepackage[english]{babel}
\usepackage{epstopdf} % Needs to come after graphicx
\usepackage{amsmath}
 \numberwithin{equation}{section}
\usepackage{mathrsfs}
\usepackage{varioref}
\usepackage{hyperref} % better to load at the last
\newcommand{\appendixpage}
\begin{document}
\setcitestyle{square}
\title{Evolution of Lifshitz metric anisotropies in Einstein-Proca theory under the Ricci-DeTurck flow %\footnote{Please insert title footnote here}%
}
%\subtitle{Do you have a subtitle?\\ If so, write it here}
\author{Roberto Cartas-Fuentevilla\inst{1},  Manuel de la Cruz\inst{1}, Alfredo Herrera-Aguilar\inst{1}, Jhony A. Herrera-Mendoza\inst{1} \and Daniel F. Higuita-Borja\inst{1}
% etc
% \thanks is optional - remove next line if not needed
\thanks{\href{mailto:rcartas@ifuap.buap.mx}{rcartas@ifuap.buap.mx}, \
\href{mailto:mdelacruz@ifbuap.buap.mx}{mdelacruz@ifuap.buap.mx},  \
\href{mailto:aherrera@ifbuap.buap.mx}{aherrera@ifuap.buap.mx}, \
\href{mailto:jhonyahm@gmail.com}{jherrera@ifuap.buap.mx}, \
\href{mailto:dfhiguit@gmail.com}{dhiguita@ifuap.buap.mx}.}%
}                     % Do not remove
%
%%\offprints{}          % Insert a name or remove this line
%
 \institute{$^1$ Instituto de F\'isica, Benem\'erita Universidad Aut\'onoma de Puebla, Apartado Postal
J-48, CP 72570, Puebla, Puebla, M\'exico.}
\date{Received: date / Revised version: date}
% The correct dates will be entered by Springer
%
\abstract{By starting from a Perelman entropy functional and considering the Ricci-DeTurck flow equations we analyze the behaviour of Einstein-Hilbert and Einstein-Proca theories with Lifshitz geometry as functions of a flow parameter. In the former case, we found one consistent fixed point that represents flat space-time as the flow parameter tends to infinity. Massive vector fields in the latter theory enrich the system under study and have the same fixed point achieved at the same rate as in the former case. The geometric flow is parametrized by the metric coefficients and represents a change in anisotropy of the geometry towards an isotropic flat space-time as the flow parameter evolves. Indeed, the flow of the Proca fields depends on certain coefficients that vanish when the flow parameter increases, rendering these fields constant. We have been able to write down the evolving Lifshitz metric solution with positive, but otherwise arbitrary, critical exponents relevant to geometries with spatially anisotropic holographic duals. We show that both the scalar curvature and matter contributions to the Ricci-DeTurck flow vanish under the flow at a fixed point consistent with flat space-time geometry. Thus, the behaviour of the scalar curvature always increases, homogenizing the geometry along the flow. Moreover, the theory under study keeps positive-definite but decreasing the entropy functional along the Ricci-DeTurck flow. 
\\
%\textbf{Keywords}: Ricci-DeTurck flow, Einstein-Proca, anisotropic space-times.
%
\PACS{
      {PACS-key}{discribing text of that key}   \and
      {PACS-key}{discribing text of that key}
     } % end of PACS codes
} %end of abstract
\maketitle

\section{Introduction and summary}
%about Ricci (DeTurck Flow)

Many physical systems have a gradient flow that codifies the dynamical behaviour of the variable of interest, for example, diffusion equations that arise in molecular dynamics and heat conduction. There are conceptual analogs of these dynamical systems with a geometric flow of manifolds with curvature. In 1982, R. S. Hamilton \cite{10.4310/jdg/1214436922} proposed the Ricci-flow equation to classify 3-dimensional manifolds in which the metric is dynamical with respect to a flow parameter
\begin{equation}
    \frac{\partial g_{\mu\nu}}{\partial\lambda}=-2 R_{\mu\nu}, \ \ \ \ g(0)=g_{0},\label{ricci1}
\end{equation}
that describes the deformation of the Riemannian metric $g_{\mu\nu}$ with respect to an \emph{auxiliary time} variable $\lambda$. This flow is analog to the heat equation in the sense that temperature homogenizes in the entire space, and corresponds to the curvature that homogenizes in all space-time.

Ricci flow equations have been used to explore a thermodynamic potential for a canonical ensemble related to the Euclidean gravity action with $S^{1}\times S^{2}$ boundary and in which the fixed points of the flow correspond to black hole stable (or unstable) configurations \cite{Headrick:2006ti}; to study the evolution, under the $\lambda$-parameter, of a two dimensional closed surface and its corresponding Hawking mass which asymptotically corresponds to the ADM energy \cite{Samuel:2007zz}; to investigate relations between Bekenstein-Hawking entropy with Perelman entropy \cite{Samuel:2007ak}; wormhole geometries, identifying different throat geometries by critical parameters, revealing topological changes on the manifold \cite{Husain:2008rg} as well as to construct maximally symmetric \cite{Cartas-Fuentevilla:2017cvt} and Lifshitz space-times \cite{Cartas-Fuentevilla:2018rez}. In the context of pure mathematics, G. Perelman made use of the Ricci flow to prove Thurston's geometrization conjecture and demonstrated the existence of a diffeomorphism map between 3-sphere into all kinds of compact $3$-manifolds without holes \cite{perelman}. Soon after Hamilton's work, D. DeTurck \cite{10.4310/jdg/1214509286} used the diffeomorphism invariance to render the Ricci flow equation into a strictly parabolic partial differential equation of the form
\begin{equation}
    \frac{\partial g_{\mu\nu}}{\partial\lambda}=-2 R_{\mu\nu}+2\nabla_{(\mu}V_{\nu )},\label{ricci2}
\end{equation}
where $V^{\mu}$ is a vector field that generates diffeomorphisms along the flow. These \emph{Ricci-DeTurck} flow equations are intriguingly similar to the so-called Holographic Renormalization Group Flow within holographic models \cite{deBoer:1999tgo}. In fact, this flow was introduced earlier by Friedan \cite{PhysRevLett.45.1057} to study the renormalization group flow for two-dimensional sigma models. The standard holographic example considers that type IIB string theory compactified on $AdS_{5}\times S^{5}$ is exactly \emph{dual} to $N$ $D3$-branes. Taking the low-energy limit, $D3$ branes yield $\mathcal{N}=4$ SYM, whereas the same low-energy limit leaves the supergravity description with $AdS_{5}\times S^{5}$. In the Poincar\'e patch of $AdS$ geometry, the radius of curvature $L$ is related to parameters of Conformal Field Theory ($SYM$) by $L=\left(g_{YM}^{2}N\right)^{1/4}$ that are described exhaustively in the \emph{boundary} of $AdS$, due to the complete match of space-time symmetries with the field theory symmetries. An important match (or \emph{dictionary}) is that the radial $r$-coordinate of bulk space-time is interpreted as a scale of the boundary field theory with a distance (energy) scale $r\sim\mu$ that controls the particle interactions.

A natural question is whether we can relate the dynamical evolution of hypersurfaces parametrized by the $r$-coordinate in bulk space-time with a Renormalization Group Flow of the theory defined on these hypersurfaces. Many attempts to relate this conceptual equivalent statements can be traced back in \cite{deBoer:1999tgo,Heemskerk:2010hk,Akhmedov:1998vf,Balasubramanian:1999jd} and a Wilsonian $\beta$-functions interpretation is given in \cite{Faulkner:2010jy}. In \cite{deBoer:1999tgo}, the RG flow for the metric is constructed using a Hamilton-Jacobi formalism with the holographic coordinate $r$ as a flow parameter. 
Despite the nature of the flow parameter, Ricci-DeTurck flow is intriguingly similar to the holographic renormalization group flow. 
In \cite{Mikula:2019tyj} the flow parameter has been implemented to describe vortex configurations in holographic descriptions of superconductivity. Besides, in this case, the flow parameter has been identified with a dimensionless Euclidean time to describe the equilibrium state in free energy functional. However, as far as the authors know, the exact relation between geometric flows\footnote{And in particular, Ricci flows.} and the RG holographic flow is yet unknown.

Concomitant with the previous paragraph, Lifshitz spaces are dual to quantum critical points in condensed matter systems \cite{Hartnoll:2016apf}. They arise in holography from a general compactification of string theory in top-down model constructions. Nowadays, holography in Lifshitz spacetimes constitutes an intensive line of research that was first considered in \cite{Kachru:2008yh}. Anisotropic space-times were also engineered within the holographic framework from a General Relativity perspective with the aid of  multiple Proca fields as well as $3$-forms with topological structure \cite{Taylor:2008tg}. Another spatially anisotropic Lifshitz metric supported by a single Proca field coupled to gravity was presented in \cite{Cartas-Fuentevilla:2018sbl} while constructing Lifshitz spaces using the Ricci flow was considered in \cite{Cartas-Fuentevilla:2018rez}. Regarding the theories\footnote{Like Massive vector field, supergravity and string theory \cite{Gregory:2010gx}.} that support a metric with anisotropic scaling symmetry $t\mapsto s^{z}t$, $x^{j}\mapsto sx^{j}$, where $z$ is the \emph{dynamical or critical exponent}, Lifshitz solutions can be considered as a generalization of AdS space-time with no boost symmetry. Then, the holographic dictionary relates these geometries with the vacuum state in the dual theory and by a suitable construction of relevant operator on the boundary, the relations between Lifshitz and AdS geometries can be investigated from the RG flow perspective. This construction is realized in \cite{Braviner:2011kz} obtaining solutions that interpolate between $AdS$ or Lifshitz geometries in the $UV$ and the $IR$ regimes. As far as the authors know,  none of these interesting papers treat the Ricci-DeTurck flow in Lifshitz spaces in the presence of matter and this work pretends to fill this gap. In order to achieve this aim, we take advantage of two facts: on the one hand, spatially anisotropic Lifshitz geometries are supported by Einstein-Proca theory with multiple vector fields, and on the other hand, these metrics are compatible with the Ricci-DeTurck flow equations.

This work is structured as follows. In Section \ref{ricciDeq} we present the Ricci-DeTurck flow equations due to matter Proca fields and their derivation from an entropy functional. In Section \ref{riccinomattermain} we present the flow for a Lifshitz metric and its fixed points in the case of Einstein-Hilbert with no matter fields. Section \ref{deformationsmatter} is devoted to the theory with multiple massive vector fields and the same Lifshitz background, the analysis of the fixed points and the behaviour of the fields and the entropy functional along the flow. Conclusions and comments are presented in final Section \ref{conclusiones}. 

\section{Ricci-DeTurck flow equations}\label{ricciDeq}

The connection between the Ricci flow and the renormalization group flow in quantum field theory has allowed to show that the former is a gradient flow defined on the space of Riemannian metrics, in such a way that an entropy functional is increasing monotonically along the flow \cite{perelman}. Based on this gradient formulation, one can construct a Ricci flow for Lorentzian metrics (or for metrics of arbitrary signature) by promoting the dynamical variables in an action principle to a family of variables connected by the corresponding Ricci flows, constituting a dynamical system defined on the space of solutions up to diffeomorphisms and gauge transformations.

Considering the Einstein-Proca action with multiple vector fields we have the family of field variables $\left(g(\lambda),\ \mathcal{A}^a(\lambda)\right)$ that define a functional on the space of metrics and connections,
\begin{equation}
  S\left(g(\lambda),\mathcal{A}^{a}(\lambda)\right)=\dfrac{1}{16\pi G_{d}} \int d^{d}x\sqrt{-g(\lambda)} \left[ R\left(g(\lambda)\right)-2\Lambda(\lambda) -\dfrac{1}{4}\sum_{a}\left[ \left( F_{a} (\mathcal{A}^{a})\right)^{2}+m_{a}^{2}(\lambda)\left(\mathcal{A}^{a}(\lambda)\right)^{2}\right]\right],
  \label{action}
\end{equation}
where $R$ is the Ricci scalar of the metric $g$, $\Lambda$ is a negative cosmological constant, $F_{a}$ are the strength tensors of the Proca fields $\mathcal{A}^{a}$ with masses $m_a$, the index $a$ labels the set of vector fields and all the field variables are functions of the flow parameter $\lambda$.

Thus, its gradient flow reads
\begin{equation}
    \delta S =\dfrac{1}{16\pi G_{d}}\int d^{d}x \sqrt{-g}\left[
    \left(
    T^{\mu\nu}-G^{\mu\nu}-\Lambda g^{\mu\nu} \right) \dfrac{\partial g_{\mu\nu}}{\partial\lambda}
    +\left( \nabla_{\mu}F_{a}^{\mu\nu}-\dfrac{1}{2}m_{a}^{2} \mathcal{A}_{a}^{\nu}\right) \dfrac{\partial \mathcal{A}^{a}_{\nu}}{\partial\lambda}
    \right] \delta\lambda,
\end{equation}
where the energy-momentum tensor is given by
\begin{equation}
    \delta\left[\left(-\dfrac{1}{4}\sqrt{-g}\right)\sum_{a}\left(F_{a}^{2}+m_{a}^{2}\mathcal{A}_{a}^{2}\right)\right]=\sqrt{-g}T^{\mu\nu}\dfrac{\partial g_{\mu\nu}}{\partial\lambda}\delta\lambda,
\end{equation}
therefore, if the evolution equations with respect to the parameter $\lambda$ are fixed to be
\begin{equation}
\dfrac{\partial g_{\mu\nu}}{\partial\lambda}+\pounds_{\widetilde{V}}g_{\mu\nu} = T_{\mu\nu}-G_{\mu\nu}-\Lambda g_{\mu\nu}; \ \ \ \ \dfrac{\partial \mathcal{A}^{a}_{\nu}}{\partial\lambda}+\pounds_{\widetilde{V}}\mathcal{A}^{a}_{\nu}=\nabla_{\mu}
    \tensor{(F^{a})}{^\mu_\nu}-\dfrac{1}{2}m_{a}^{2}\mathcal{A}^{a}_{\nu},\label{riccidt0}
\end{equation}
where $\pounds_{\widetilde{V}}g_{\mu\nu}$ and $\pounds_{\widetilde{V}}\mathcal{A}^{a}_{\mu}$ are modifying the flows by a diffeomorphism generated by the vector field $\widetilde{V}$
\begin{equation}
\pounds_{\widetilde{V}} g_{\mu\nu} = 2 \nabla_{(\mu}\widetilde{V}_{\nu)}, \ \ \ \
         \pounds_{\widetilde{V}}\mathcal{A}^{a}_{\mu} = \widetilde{V}^{\nu}\partial_{\nu}\mathcal{A}^{a}_{\mu}+\mathcal{A}^{a}_{\nu}\partial_{\mu}\widetilde{V}^{\nu}, \ \ \ \
         \pounds_{\widetilde{V}}\mathcal{A}_{a}^{\mu} = \widetilde{V}^{\nu}\partial_{\nu}\mathcal{A}_{a}^{\mu}-\mathcal{A}_{a}^{\nu}\partial_{\nu}\widetilde{V}^{\mu},
\end{equation}
then, the gradient flow reduces to
\begin{equation}
    \delta S =\dfrac{1}{16\pi G_{d}}\int d^{d}x \sqrt{-g}\left[ \left( T_{\mu\nu}-G_{\mu\nu}-\Lambda g_{\mu\nu}\right)^{2}+\left( \nabla_{\mu}F_{a}^{\mu\nu}-\dfrac{1}{2}m_{a}^{2}\mathcal{A}^{\nu}\right)^{2}  \right]\delta\lambda.\label{entropy}
\end{equation}

Thus, for Riemannian metrics, this entropy functional is increasing along the flows, however, for a general metric with arbitrary signature, the functional may decrease and/or increase along the flows\footnote{See for example \cite{Woolgar:2007vz} for some applications of Ricci flows in General Relativity.}.

The \emph{fixed points} of the dynamical system are those solutions that do not change in $\lambda$-\emph{time}; thus, they are geometries for which
\begin{equation}
    \partial_{\lambda}g_{\mu\nu}=0=T_{\mu\nu}-G_{\mu\nu}-\Lambda g_{\mu\nu}, \ \ \ \ \partial_{\lambda}\mathcal{A}^{a}_{\nu}=0=\nabla_{\mu}\left(F_{a}\right)^{\mu\nu}-\dfrac{1}{2}m_{a}^{2}\mathcal{A}^{\nu};
\end{equation}
these geometries correspond to solutions of equations of motion related with the theory (\ref{action}). 

In this work, we consider the addition of matter fields by means of the \emph{reduced} energy-momentum tensor of the Einstein-Proca theory defined as
\begin{equation}
    \bar{T}_{\mu\nu}=\dfrac{2\Lambda}{d-2}g_{\mu\nu}+\dfrac{1}{4}\sum_{a}\left[2\left(F^{a}\right)_{\mu\rho}\tensor{\left(F^{a}\right)}{_\nu^\rho}+m_{a}^{2}\mathcal{A}^{a}_{\mu}\mathcal{A}^{a}_{\nu}-\dfrac{1}{d-2}\left(F^{a}\right)^{2}g_{\mu\nu}\right],\label{EMtensor}
\end{equation}
in such a way that the flow of Ricci-DeTurck equations (\ref{riccidt0}) evolve as follows
\begin{subequations}\label{RicciDTmain}
\begin{align}
   \partial_{\lambda}g_{\mu\nu} &=-2 R_{\mu\nu}+2\bar{T}_{\mu\nu}+2\nabla_{(\nu}V_{\mu)}, \qquad \qquad V_{\mu}=-2\widetilde{V}_{\mu}; \label{RicciDT1}\\
   \partial_{\lambda}\mathcal{A}^{a}_{\nu} &=2\nabla^{\mu}F_{\mu\nu}-m_{a}^{2}\mathcal{A}^{a}_{\nu}+V^{\mu}\nabla_{\mu}\mathcal{A}^{a}_{\nu}+\mathcal{A}^{a}_{\mu}\nabla_{\nu}V^{\mu},\label{RicciDT2}
\end{align}
\end{subequations}
where we have accordingly rescaled the flow parameter when passing from Eq. (\ref{riccidt0}) to Eq. (\ref{RicciDTmain}).
We call these equations \emph{Ricci-DeTurck} flow equations and $V^{\mu}$ is the DeTurck vector field associated with diffeomorphisms along the flow. 

The equations (\ref{RicciDTmain}) constitute the principal framework of this work. To begin their analysis, we consider that the action (\ref{action}) depends on $\lambda$ through the dynamical fields and the following metric 
\begin{equation}
ds^{2}=dr^{2}+\sum_{i=0}^{d-2}e^{2\gamma_{i}r}\eta_{ij}dx^{i}dx^{j}, \ \ \  \ i,j=0,...,d-2,\label{metric1}
\end{equation}
has a set of arbitrary $\gamma_{i}(\lambda)$ coefficients that encode both the space-time and purely spatial anisotropic structure of the Lifshitz geometry. The metric $\eta_{ij}$ stands for the Minkowski space-time. 

On the other hand, the ansatz for the vector fields is taken to be 
\begin{equation}
\mathcal{A}^{a}_{k}=A^{a}_{k}e^{\beta_{k}r}, \ \ \ \ a=0,...d-2,\label{vf}
\end{equation}
where the coefficients $A^{a}_{k}$ are considered to be constant and $\beta_{k}(\lambda)$, appearing in the argument of the exponential, constitute another set of coefficients that need to be determined. The DeTurck vector field $V^{\mu}(r,\lambda)$ depends on both the flow parameter and the holographic coordinate for consistency.
Throughout this paper we consider metrics with $\left( -,\textbf{+}\right)$ signature; Greek indices run over $\left(0,d-1\right)$ and Latin indices run over $\left(0,d-2\right)$. The vector field ansatz (\ref{vf}) is compatible with the metric (\ref{metric1}) since the $r$-component of the latter has no exponential factor and we shall consider that the $r$-components of the Proca fields $A^{a}_{r}$ vanish for all indices $a$. 

Finally, we would like to stress that the $\gamma_{i}$ parameters of the metric, the cosmological constant, the $\beta_{k}$ coefficients of the vector fields and their masses $m_a$ are assumed to have all $\lambda$ functional dependency. In what follows we shall explore the Ricci-DeTurck flow (\ref{RicciDTmain}) for the Einstein-Proca theory (\ref{action}).

\section{Geometric Ricci-DeTurck flow}\label{riccinomattermain}

Before proceeding to analyze the flow equations of the complete Einstein-Proca theory, it is important to obtain qualitative and quantitative insights into the behaviour of the pure Lifshitz geometry (\ref{metric1}) under the Ricci-DeTurck flow, i.e., without considering any matter fields as we shall see.

Under the geometric Ricci-DeTurck flow (\ref{ricci2}), we have
\begin{equation}
    \begin{array}{rcl}
         \partial_{\lambda}g_{rr} &=& -2R_{rr}+2\nabla_{(r}V_{r)}\\
         \partial_{\lambda}g_{ij} &=& -2R_{ij}+2\nabla_{(i}V_{j)}. \label{flownomatter}
    \end{array}
\end{equation}

Considering the metric (\ref{metric1}), the components of the Ricci tensor are
\begin{equation}
R_{rr}=-\sum_{k=0}^{d-2}\gamma_{k}^{2}(\lambda), \ \ \ \ \ R_{ij}=-\eta_{ij}e^{2\gamma_{i}(\lambda)r}\gamma_{i}(\lambda)\sum_{k=0}^{d-2}\gamma_{k}(\lambda).\label{riccigeom}
\end{equation}
Then, Eqs. (\ref{flownomatter}) read (suppressing the $\lambda$-dependency for simplicity)
\begin{subequations}\label{riccinomatter1}
\begin{align}
  0 &= \sum_{k=0}^{d-2}\gamma_{k}^{2}+\nabla_{r}V_{r},\\ 
  r\dot{\gamma}_{i}\eta_{ij} &= \eta_{ij}\gamma_{i}\sum_{k=0}^{d-2}\gamma_{k}+e^{-2\gamma_{i}r}\nabla_{(i}V_{j)},
\end{align}
\end{subequations}
where an overdot denotes derivatives with respect to the flow parameter $\lambda$. 
The structure of the latter equations implies that the DeTurck vector must have the form
\begin{equation}
    V_{r}=-A(\lambda)r+B(\lambda), \label{VAB}
\end{equation}
and is the unique $(r, \lambda)$ functional dependency consistent with them. Therefore, from Eqs. (\ref{riccinomatter1}) we get
\begin{subequations}\label{riccinomatter2}
\begin{align}
    0 &= \sum_{k=0}^{d-2}\gamma_{k}^{2}-A,\\
    r\dot{\gamma}_{i} &= \gamma_{i}\sum_{k=0}^{d-2}\gamma_{k}+\gamma_{i}\left(-rA+B\right), \ \ \ \forall \ i=0,...,d-2;
\end{align}
\end{subequations}
which in turn render the functions $A(\lambda)$ and $B(\lambda)$ in terms of $\gamma_{k}$ as well as a differential equation for these coefficients
\begin{subequations}\label{riccinomatter3}
\begin{align}
    A(\lambda) &= \sum_{k=0}^{d-2}\gamma_{k}^{2},\label{rnm1}\\
    B(\lambda) &= -\sum_{k=0}^{d-2}\gamma_{k},\label{rnm2}\\
    \dot{\gamma}_{i} +\gamma_{i}A(\lambda) &= 0, \ \ \ \forall \ i=0,...,d-2.\label{rnm3}
\end{align}
\end{subequations}
If we take two arbitrary $\gamma_{i}$ in (\ref{rnm3}) with different index, we can see that $\gamma_{i}$ and $\gamma_{j}$ are proportional to each other, for all $i$ and $j$. Therefore, we can write  
\begin{equation}
    \gamma_{i}(\lambda)\equiv \chi(\lambda) z_{i}, \ \ \ z_{i}>0 \ \ \forall \ i, \label{criticalexpo}
\end{equation}
where the proportionality factors $z_{i}$ are associated with the anisotropic structure of the metric (\ref{metric1}), representing critical exponents\footnote{We mentioned in the introduction section that, within the holographic framework, the so-called \emph{critical exponents} $z_{i}$ of a Lifshitz geometry are to be related with order parameters in quantum critical points of a dual field theory.} that can be considered positive, but otherwise arbitrary. Therefore, the equation (\ref{rnm3}) now reads
\begin{equation}
\dot{\chi}(\lambda)=-\chi^{3}(\lambda)\sum_{k=0}z_{i}^{2},
\end{equation}
with solution
\begin{equation}
    \chi(\lambda) =%\pm 
    \left[\dfrac{1}{2\sum_{k=0}^{d-2}z_{k}^{2} \left(\lambda-\lambda_{0}\right)}\right]^{1/2}= %\pm 
    \left[\dfrac{1}{2Z \left(\lambda-\lambda_{0}\right)}\right]^{1/2}, \ \ \ Z\equiv \sum_{k=0}^{d-2}z_{k}^{2},\label{lasjis0}
\end{equation}
where we chose the $(+)$ sing in $\chi(\lambda)$ in order to be consistent with the positive signs of the $z_i$ critical exponents and the defined $Z$ constant will be useful in forthcoming sections. We therefore have now the functions $A$ and $B$ that define the DeTurck vector field $V_{\mu}$ according to (\ref{VAB}) completely determined by the critical exponents $z_{i}$ and the flow parameter $\lambda$. The explicit expressions for $ A(\lambda)$ and $B(\lambda)$ as functions of the flow parameter adopt the form
\begin{equation}
\begin{array}{rcl}

A(\lambda) &=& \chi^{2}\sum_{k=0}^{d-2}z_{k}^{2} = %\dfrac{\sum_{k=0}^{d-2}z_{k}^{2}}{2Z\left(\lambda-\lambda_{0}\right)}=
\dfrac{1}{2\left(\lambda-\lambda_{0}\right)}, \\ \\

B(\lambda) &=& -\sum_{k=0}^{d-2}\gamma_{k} = -\dfrac{\sum_{k=0}^{d-2}z_{k}}{\left[2Z \left(\lambda-\lambda_{0}\right)\right]^{1/2}}.
\label{ABsolutions}
\end{array}
\end{equation}

As a result of this analysis, the metric (\ref{metric1}) is now an evolving metric due to $\lambda$ dependency
\begin{equation}
    ds^{2}=dr^{2}+\sum_{i=0}^{d-2}e^{2z_{i}\chi(\lambda)r}\eta_{ij}dx^{i}dx^{j},\label{metric1.1}
\end{equation}
and has a \emph{fixed point} in the limit $\lambda\mapsto\infty$, which is the flat space-time on $d$-dimensions
\begin{equation}
    ds^{2}_{\lambda\mapsto\infty}=dr^{2}+\eta_{ij}dx^{i}dx^{j}.
\end{equation}
This implies that the curvature scalar must vanish in that limit and, indeed, it does:
\begin{equation}
\begin{array}{rcl}
R = -\sum_{k=0}^{d-2}\gamma_{k}^{2}-\left( \sum_{k=0}^{d-2}\gamma_{k}\right)^2
    =-\frac{1}{2Z\left(\lambda-\lambda_{0}\right)} \left[ \sum_{k=0}^{d-2}z_{k}^{2}+\left( \sum_{k=0}^{d-2}z_{k}\right)^2
    \right];
    \label{ricciscalar}
    \end{array}
\end{equation}
thus, the negative-definite Ricci scalar tends to zero as the flow goes by. 

%\subsection{Fixed points of the flow}

General fixed points in the Ricci-DeTurck flow are solutions of the equation
\begin{equation}
    \partial_{\lambda}g_{\mu\nu}=0;\label{fixedpointdef}
\end{equation}
from the evolving metric (\ref{metric1.1}) fixed point equations read
\begin{equation}
\begin{array}{rcl}
\partial_{\lambda}g_{rr} &=& 0,\\ \\

\partial_{\lambda}g_{ij} &=& - 2r\eta_{ij}\dfrac{ Z z_{i}}{\left[ 2Z\left(\lambda-\lambda_{0}\right)\right]^{3/2}}e^{2\chi z_{i}r}=0,\\ 
\end{array}\label{fixedpnomatter}
\end{equation}
ensuring that the fixed point of anisotropic Lifshitz geometries corresponds to flat space-time under the evolution of the Ricci-DeTurck flow parameter.

Thus, we have shown that the pure geometric Ricci-DeTurck flows compatible with a completely anisotropic Lifshitz space, where the metric 
coefficients are parameterized by distinct critical exponents, have Minkowski space-time as a fixed point under the flow evolution. In other words,
by starting with a totally anisotropic (spatially as well) geometry with constant negative Ricci curvature, as the flow parameter $\lambda$ evolves the metric 
isotropizes and tends to flat space-time while the scalar curvature vanishes in accordance with the monotonic increasing behaviour predicted by a 
geometric flow.

\section{Ricci-DeTurck flow in the presence of Proca fields}\label{deformationsmatter}

In this section we consider the Ricci-DeTurck flow for the full Einstein-Proca theory, i.e. in the presence of massive vector fields with the relevant flow equations (\ref{RicciDTmain}). 

The \emph{reduced} stress-energy tensor (\ref{EMtensor}) for the Einstein-Proca theory (\ref{action}) has the form (suppressing the $\lambda$-dependency for simplicity)
\begin{equation}
	\begin{array}{rcl}
		\bar{T}_{rr} &=& \dfrac{2\,\Lambda}{d-2}+ \dfrac{d-3}{2(d-2)}\sum_{a}\sum_{k=0}e^{2\left(\beta_{k}-\gamma_{k}\right)r}\eta^{kk}\beta_{k}^2\,{A^{a}_{k}}^2,\\ \\
		\bar{T}_{ij} &=& \dfrac{2\, \Lambda}{d-2}\,\eta_{ij}\, e^{2\gamma_{i}r}+\dfrac{1}{4}\sum_{a}\left[\Bigl(2\beta_{i}\beta_{j}+m_{a}^{2}\Bigr)A^{a}_{i}A^{a}_{j}e^{\left(\beta_{i}+\beta_{j}\right)r}-\dfrac{2}{d-2}\eta_{ij}\, e^{2\gamma_{i}r}\sum_{k=0} e^{2\left(\beta_{k}-\gamma_{k}\right)r} \eta^{kk} \beta_{k}^2 {A^{a}_{k}}^2 \right].
		\label{stressen}
	\end{array}
\end{equation}

Armed with the expressions for the components of the Ricci tensor (\ref{riccigeom}) and the stress-energy tensor (\ref{stressen}), the Ricci-DeTurck flow equations for both the metric and the Proca fields (\ref{RicciDTmain}) read
\begin{itemize}
	\item for the $rr$ component of the metric flow (\ref{RicciDT1}) 
\end{itemize}
\begin{equation}\label{eq:rr}
	0 = \sum_{k=0}\gamma_{k}^{2}+\,\partial_{r}V_{r}+\dfrac{2\,\Lambda}{d-2}+ \dfrac{d-3}{2(d-2)}\sum_{a}\sum_{k=0}e^{2\left(\beta_{k}-\gamma_{k}\right)r}\eta^{kk}\beta_{k}^2\,{A^{a}_{k}}^2.
\end{equation}
\begin{itemize}
	\item for the $ij$-component of the metric flow (\ref{RicciDT1}) 
\end{itemize}
\begin{equation}
	\begin{split}
		r\,\dot{\gamma}_{i}\,\eta_{ij} =& \dfrac{2\, \Lambda}{d-2}\,\eta_{ij}+\dfrac{1}{4}\sum_{a}\left[\Bigl(2\beta_{i}\beta_{j}+m_{a}^{2}\Bigr)A^{a}_{i}A^{a}_{j}e^{\left(\beta_{i}+\beta_{j}-2\gamma_{i}\right)r}-\dfrac{2}{d-2}\eta_{ij}\, \sum_{k=0} e^{2\left(\beta_{k}-\gamma_{k}\right)r} \eta^{kk} \beta_{k}^2 {A^{a}_{k}}^2 \right] \label{eq:ij}\\
		&+\gamma_{i}\, \eta_{ij}\Bigl(\sum_{k=0}\gamma_{k}+ V_{r}\Bigr).
	\end{split}
\end{equation}
\begin{itemize}
	\item for the $i$-component of the Proca vector field flow (\ref{RicciDT2}) 
\end{itemize}
\begin{equation}
r\dot{\beta_i}{A}^{a}_{i} =\left(2\beta_i^2 + 2\sum_{k=0}^{d-2}\gamma_{k}\beta_i-4\gamma_i\beta_i-m_{a}^{2}+\beta_i\,V_r\right) {A}^{a}_{i}\,,
    \label{procaflow2}
\end{equation}
since the $r$-component gives no contribution to this flow.

Before proceeding with our construction, it is worthwhile to consider the off-diagonal elements of equation (\ref{eq:ij}) implying the following restriction
\begin{equation}
	\dfrac{1}{4}\sum_{a}\left( 2\beta_{i}\beta_{j}+m_{a}^{2}\right)A^{a}_{i}A^{a}_{j}=0, \qquad  i\neq j.
	\label{restriction}
\end{equation}

On the other hand, by writing the equation (\ref{procaflow2}) in matrix form 
one realizes that it includes two cases. In the first one, this equation can be satisfied if and only if the masses of the Proca vector fields are identical $m_{a}^{2}=m^2$ for all $a$ values, a fact that implies in turn that $\beta_{i}=\beta$ for all $i$, yielding a single effective Proca field that is not able to support a completely spatial anisotropic Lifshitz geometry \cite{Cartas-Fuentevilla:2018sbl}. The second case consists in considering different masses for the Proca fields, forcing the vector fields to have the following form
\begin{equation}\label{eq:Proca_cst}
	\mathcal{A}^{a}_{k}=A_{k}\delta^{a}_{k}e^{\beta_{k}r},
\end{equation} 
which also satisfies Eq. (\ref{restriction}). The structure of the vector field (\ref{eq:Proca_cst}) was proposed in \cite{Taylor:2008tg} when constructing a partial spatially anisotropic geometry supported by Lifshitz gravity coupled to several Proca fields. The latter case is the one we will deal with from here on, from which Eqs. \eqref{eq:rr}-\eqref{procaflow2} now read
\begin{subequations}
	\begin{align}
		0 =& \sum_{k=0}\gamma_{k}^{2}+\partial_{r}V_{r}
		+\dfrac{2\,\Lambda}{d-2}+\dfrac{(d-3)}{2(d-2)}\sum_{k=0}e^{2\left(\beta_{k}-\gamma_{k}\right)r}\eta^{kk}\, \beta_{k}^2 \,A_{k}^2\,, \label{eq:rr_2}\\ 
		r\,\dot{\gamma}_{i}\,\eta_{ij} =&\ \gamma_{i}\, \eta_{ij}\Bigl(\sum_{k=0}\gamma_{k}+V_{r}\Bigr)+ \dfrac{2\, \Lambda}{d-2}\eta_{ij}+\dfrac{1}{4}\Bigl(2\beta_{i}^2+m_{i}^2\Bigr)\,\delta_{ij}A_{i}^2\, e^{2\left(\beta_{i}-\gamma_{i}\right)r}\label{eq:ij_2}\\
		&- \dfrac{\eta_{ij}}{2(d-2)}\, \sum_{k=0}e^{2\left(\beta_{k}-\gamma_{k}\right)r} \eta^{kk}\,\beta_{k}^2\, A_{k}^2\,, \nonumber\\
r\dot{\beta_i}=&2\beta_i^2 + 2\sum_{k=0}^{d-2}\gamma_{k}\beta_i-4\gamma_i\beta_i-m_{i}^{2}+\beta_i\,V_r \,.
    \label{eq:Ai_2}
	\end{align}
\end{subequations}

The structure of these equations leads to the same functional dependency for the DeTurck vector field\footnote{In principle one can integrate Eq. (\ref{eq:rr_2}) and get  linear in $r$ expression supplemented by an exponential term for the DeTurck vector field. However, there is no way to get rid of this term in Eq. (\ref{eq:Ai_2}) without trivializing the vector field. Hence, this case reduces to the one treated in Sec. \ref{riccinomattermain}.}
\begin{equation}
	V_{r}=-g(\lambda)\,r+f(\lambda),\label{ansatzV2}
\end{equation}
as in the previous Section. Hence, Eqs. \eqref{eq:rr_2}-\eqref{eq:Ai_2} adopt the form

\begin{subequations}
	\begin{align}
		0 =& \sum_{k=0}\gamma_{k}^{2}-g(\lambda)
		+ \dfrac{2\,\Lambda}{d-2}+\dfrac{(d-3)}{2(d-2)}\sum_{j,k=0}e^{2\left(\beta_{k}-\gamma_{k}\right)r}\eta^{kk}\, \beta_{k}^2 \,A_{k}^2\,,\label{eq:rr_3}\\ 
		r\,\dot{\gamma}_{i}\,\eta_{ij} =&\ \gamma_{i}\, \eta_{ij}\Bigl[\sum_{k=0}\gamma_{k}-g(\lambda)\,r+f(\lambda) \Bigr]+\dfrac{2\, \Lambda}{d-2}\eta_{ij}+\dfrac{1}{4}\Bigl(2\beta_{i}^2+m_{i}^2\Bigr)\,\delta_{ij}A_{i}^2\, e^{2\left(\beta_{i}-\gamma_{i}\right)r}\,\label{eq:ij_3}\\
		&- \dfrac{\eta_{ij}}{2(d-2)}\, \sum_{k=0}e^{2\left(\beta_{k}-\gamma_{k}\right)r} \eta^{kk}\,\beta_{k}^2\, A_{k}^2\,,\nonumber\\
r\dot{\beta_i}=&2\beta_i^2 + 2\sum_{k=0}^{d-2}\gamma_{k}\beta_i-4\gamma_i\beta_i-m_{i}^{2}+\beta_i\,\Bigl[-g(\lambda)\,r+f(\lambda) \Bigr] \,.
    \label{eq:Ai_3}
	\end{align}
\end{subequations}

The consistency of Eq. \eqref{eq:rr_3} forces us to set
\begin{equation}
	\beta_{i}=\gamma_{i},\ \qquad \ \forall \ i=0,...,d-2. \label{betagama}
\end{equation}
which yields the following system of equations
\begin{subequations}\label{restritcall}
	\begin{align}
		g(\lambda) &=\sum_{k=0}\gamma_{k}^{2}+\dfrac{2\Lambda}{d-2}+\dfrac{d-3}{2(d-2)}\sum_{k=0}\eta^{kk}\,\gamma_{k}^2\, A_{k}^2\,,\label{restrict1}\\ 
		-\gamma_{i}\,f(\lambda)\,\eta_{ij} &=\sum_{k=0}\gamma_{k}\, \gamma_{i}\,\eta_{ij}+\dfrac{2\Lambda}{d-2}\eta_{ij} + \dfrac{1}{4}\Bigl(2\gamma_{i}^2+m_{i}^2\Bigr)\, A_{i}^2\,\delta_{ij}
		-\dfrac{1}{2(d-2)}\sum_{k=0}\eta^{kk}\, \gamma_{k}^2\, A_{k}^2\,\eta_{ij}\,,\label{restrict2} \\
		0 &= \dot{\gamma}_{i}+\gamma_{i}\,g(\lambda)\,, \label{restrict3}\\
m_{i}^{2}=& 2\sum_{k=0}^{d-2}\gamma_{k}\,\gamma_i-2\gamma_i^2+f(\lambda)\,\gamma_i\,.
    \label{restrict4}
	\end{align}
\end{subequations}
from which we need to determine $\lbrace \Lambda, \gamma_{i}, f,g, m_{i}\rbrace$ in terms of the flow parameter $\lambda$ and the $A_{k}$ vector coefficients, a task that will be worked out next.
%%%%%%%%%%%%%%%%%%%%%%%%%%%%%%%%%%%%%%%%%%%%%%%%%%%%%%%%%%%%%%%%%%%%%%%%%%%%%%%%%
%%%%%%%%%%%%%%%%%%%%%%%%%%%%%%%%%%%%%%%%%%%%%%%%%%%%%%%%%%%%%%%%%%%%%%%%%%%%%%%%%
%%%%%%%%%%%%%%%%%%%%%%%%%%%%%%%%%%%%%%%%%%%%%%%%%%%%%%%%%%%%%%%%%%%%%%%%%%%%%%%%%
\subsection{Solving the system}\label{s3}

In order to solve the system $\lbrace \Lambda, \gamma_{i}, f,g, m_{i}\rbrace$ it is convenient to start separating the temporal and spatial parts in  (\ref{restrict2})
\begin{subequations}\label{temporalspatial}
   \begin{align}
   \gamma_{0}\left(1-\dfrac{1}{4}A_{0}^{2}\right)f(\lambda) &= -\sum_{k=0}\gamma_{k}\gamma_{0}-\dfrac{2\Lambda}{d-2}+\dfrac{1}{2}\sum_{k=0}\gamma_{k}\gamma_{0}A_{0}^{2}+\dfrac{1}{2(d-2)}\sum_{k=0}\eta^{kk}\gamma_{k}^{2}A_{k}^{2}, \label{temporalspatial1}\\
   -\sum_{l=1}\gamma_{l}\left(1+\dfrac{1}{4}A_{l}^{2}\right)f(\lambda) &= \sum_{k=0}\gamma_{k}\sum_{l=1}\gamma_{l}+2\Lambda+\dfrac{1}{2}\sum_{k=0}\gamma_{k}\sum_{l=1}\gamma_{l}A_{l}^{2}-\dfrac{1}{2}\sum_{k=0}\eta^{kk}\gamma_{k}^{2}A_{k}^{2}.\label{temporalspatial2}
   \end{align}
\end{subequations}
where we have used the Eq. (\ref{restrict4}) to eliminate the dependency of $m_{i}$ and sum the spatial part (\ref{restrict2}) over the spatial coordinates. Multiplying by $1/(d-2)$ this spatial part and summing with (\ref{temporalspatial1}), we obtain an expression for the function $f(\lambda)$, namely
\begin{equation}
f(\lambda)= -\sum_{k=0}\gamma_{k}\left[1+h\left(A_{k},\gamma_{k}\right)\right],\label{fsolved1}
\end{equation}
where we have defined
\begin{equation}
h\left(A_{k},\gamma_{k}\right)\equiv \dfrac{(d-3)A_{0}^{2}\gamma_{0}+\sum_{k=0}A_{k}^{2}\gamma_{k}}{-4(d-1)\gamma_{0}+4\sum_{k=0}\gamma_{k}+(d-3)A_{0}^{2}\gamma_{0}+\sum_{k=0}A_{k}^{2}\gamma_{k}},\label{fsolved2}
\end{equation}
and the denominator of $h\left(A_{k},\gamma_{k}\right)$ is restricted to be non-zero.

It is important to stress that the case $A_{k}=0$ for all $k$, reduces the Eq. (\ref{fsolved1}) to  the $B(\lambda)$ function in the purely geometric Ricci-DeTurck flow Eq. (\ref{ABsolutions}).

Taking the temporal component (\ref{temporalspatial1}) and considering the function (\ref{fsolved1}), an expression for the cosmological constant is obtained
\begin{equation}
\dfrac{2\Lambda}{d-2}=\dfrac{1}{4}\sum_{k=0}\gamma_{k}\gamma_{0}\left[A_{0}^{2}+4h-A_{0}^{2}h\right]+\dfrac{1}{2(d-2)}\sum_{k=0}\eta^{kk}A_{k}^{2}\gamma_{k}^{2},\label{Cosmo}
\end{equation}
whereas, from the spatial components of (\ref{restrict2}) with no sum, we obtain $d-2$ restrictions between $\gamma_{k}$ and $A_{k}$:
\begin{equation}
4h%\left(A_{k},\gamma_{k}\right)
\left(\gamma_{k}-\gamma_{0}\right)+\left(\gamma_{0}A_{0}^{2}+\gamma_{k}A_{k}^{2}\right)\left(h
%\left(A_{k},\gamma_{k}\right)
-1\right)=0,\label{restriction5}
\end{equation}
which can always be satisfied.

On the other hand, each mass of the vector fields (\ref{restrict4}) adopts the form
\begin{equation}
m_{i}^{2}=\sum_{k=0}\gamma_{k}\left(1-h\right)\gamma_{i}-2\gamma_{i}^{2}.\label{masas}
\end{equation}

Also, with the help of (\ref{Cosmo}), the $g(\lambda)$ function in (\ref{restrict1}) reads
\begin{equation}
g(\lambda)=\sum_{k=0}\gamma_{k}^{2}+\dfrac{1}{4}\sum_{k=0}\gamma_{k}\gamma_{0}\left[A_{0}^{2}+4h-A_{0}^{2}h\right]+\dfrac{1}{2}\sum_{k=0}\eta^{kk}\gamma_{k}^{2}A_{k}^{2}.\label{gfunction}
\end{equation}
Having obtained the functions $\lbrace \Lambda, f,g, m_{i}\rbrace$ we can solve for the metric coefficients $\gamma_{i}$. By analyzing the structure of Eq. (\ref{restrict3}) where the $\gamma_{i}$ coefficients depend on $\lambda$, we are led to the same result as in Sec. \ref{riccinomattermain}:
\begin{equation}
	\gamma_{i}\equiv z_{i}\chi(\lambda), \ \ \forall \ i,\label{chidef}
\end{equation}
with arbitrary critical exponents $z_i$. From this relation, Eq. (\ref{restrict3}) renders
\begin{equation}
	\dot{\chi}=-\chi g(\lambda),\label{solving}
\end{equation}
where $g(\lambda)$ is now known according to Eq. (\ref{gfunction}). Because of the relation (\ref{chidef}), the $h$ function becomes independent of $\lambda$.

\begin{figure}[H]
\begin{center}
 \includegraphics[width=0.6\textwidth]{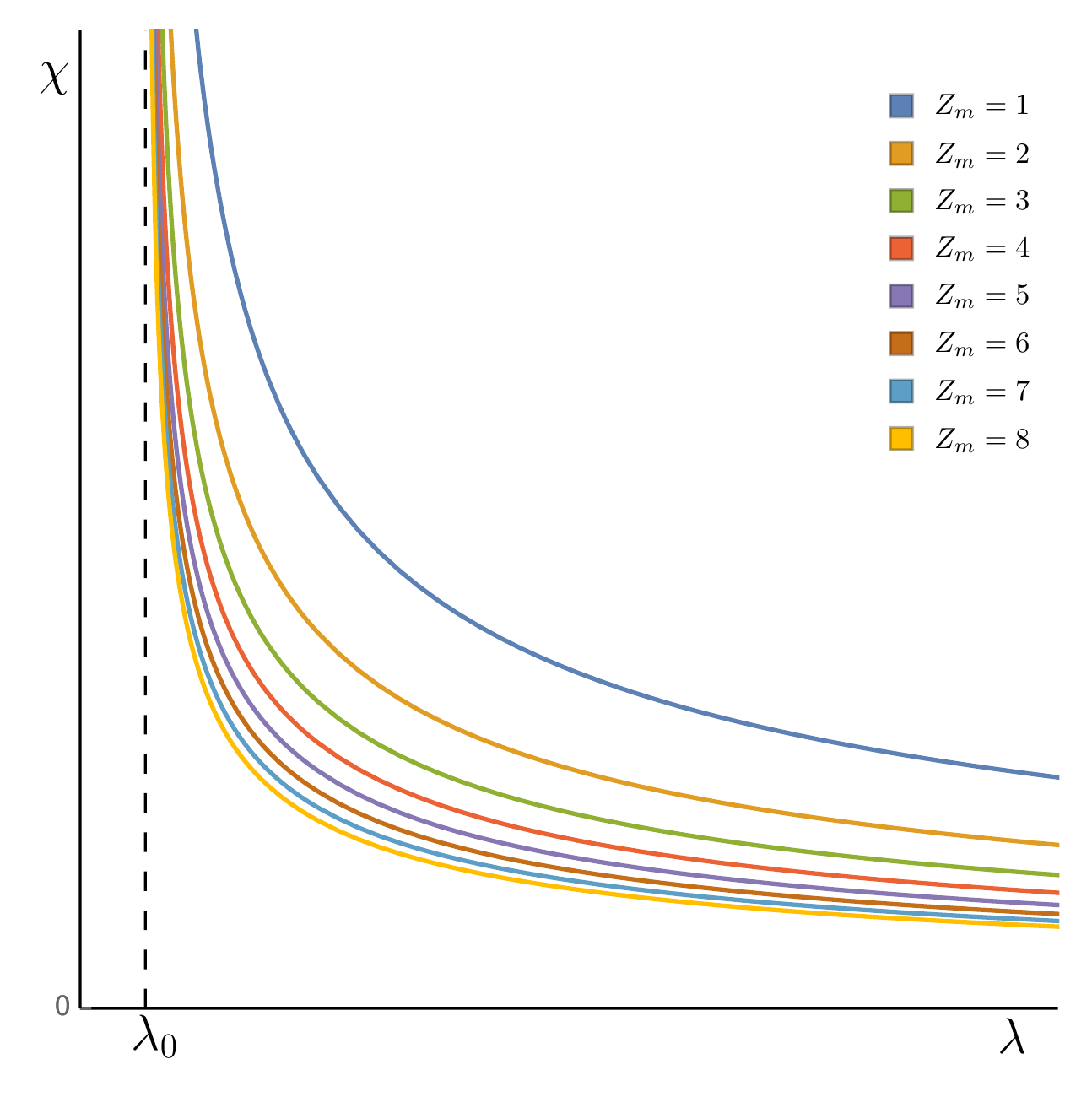}
 % A.png: 427x418 pixel, 72dpi, 15.06x14.75 cm, bb=0 0 427 418
 \caption{Behavior of $\chi(\lambda)$ for different values of $Z_{m}>0$ and initial flow condition $\lambda_{0}$. For all cases, there is a $\chi\mapsto 0$ convergence.}
 \label{fig:1}
 \end{center}
\end{figure}

On the other hand, analogue to the case of pure geometric flow (\ref{riccinomattermain}), we define 
\begin{equation}
Z_{m}\equiv \sum_{k=0}z_{k}^{2}+\dfrac{1}{4}\sum_{k=0}z_{k}z_{0}\left[A_{0}^{2}+4h-A_{0}^{2}h\right]+\dfrac{1}{2}\sum_{k=0}\eta^{kk}A_{k}^{2}z_{k}^{2},
\end{equation}
which reduces to pure geometric Ricci-DeTurck flow when all the vector fields and their masses vanish (see the last equation in (\ref{lasjis0})). From this definition, the function $\chi(\lambda)$ in (\ref{solving}) with $\lambda_{0}$ as initial flow condition, is completely integrable
\begin{equation}
    \chi(\lambda)=\left[ \frac{1}{2Z_{m} \left(\lambda-\lambda_{0}\right) }  \right]^{1/2}.\label{lasjis}
\end{equation}
Analyzing this solution, the positiveness (or negativeness) of $Z_{m}$ can be ensured with a suitable choice of the coefficients of the Proca fields and the anisotropic critical exponents. For simplicity and without loss of generality, it is lawful to assume that $Z_{m}$ is positive.
Recall that from the physical standpoint, we assumed that all the $\gamma_{i}>0$ in order to relate them with the positive critical exponents for possible dual Condensed Matter Theories (see \cite{Cartas-Fuentevilla:2018sbl} for the case of Lifshitz geometries). Taking these considerations together, we choose $\chi >0$ in (\ref{lasjis}) from now on. In Figure (\ref{fig:1}) we show the behaviour of the $\chi(\lambda)$ function under the evolution of the flow parameter for different values of $Z_{m}$.  

\subsection{Evolution of the system under $\lambda$ and fixed points}

Having obtained the explicit expression for the function $\chi(\lambda)$, the metric (\ref{metric1}) and the Proca fields (\ref{eq:Proca_cst}) become evolving functions under the flow parameter
\begin{equation}
ds^{2}=dr^{2}+\sum_{i=0}^{d-2}e^{2\chi(\lambda) z_{i}r}\eta_{ij}dx^{i}dx^{j}\,, \qquad \mathcal{A}_k^a=A_k\delta_k^ae^{\chi(\lambda) z_{k}r}\,.
\label{metric2}
\end{equation}
We see that the metric and the Proca fields in (\ref{metric2}) are well defined in the interval $\lambda_{0} < \lambda < \infty$ due to the behaviour of (\ref{lasjis}). 

\begin{figure}[H]
\begin{center}
 \includegraphics[width=0.6\textwidth]{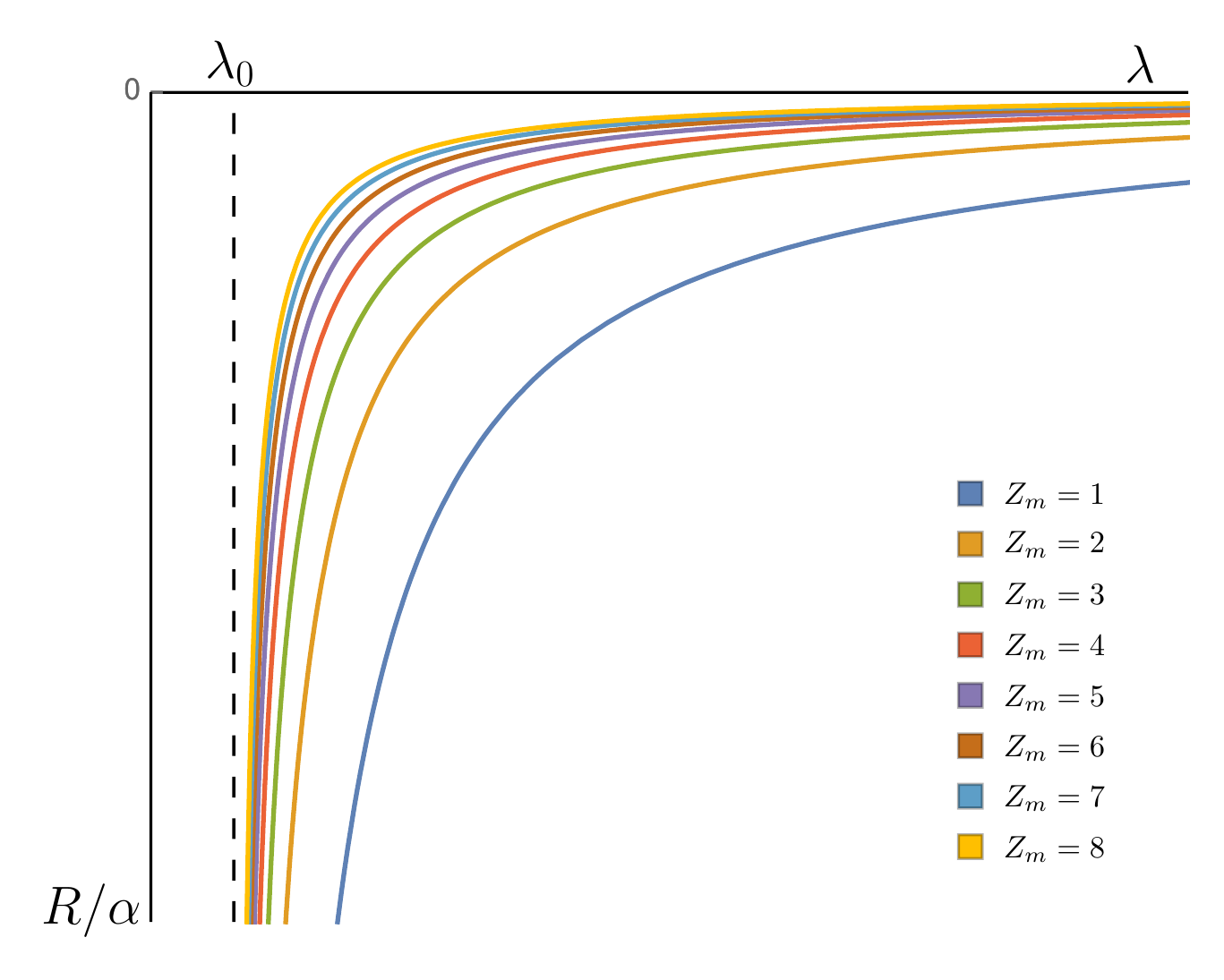}
 % A.png: 427x418 pixel, 72dpi, 15.06x14.75 cm, bb=0 0 427 418
 \caption{Behavior of $R/\alpha$ where $\alpha$ is the numerator in (\ref{Scalarflow}). For different values of $Z_{m}>0$ and initial flow condition $\lambda_{0}$, there is a $R\mapsto 0$ flat convergence.}
 \label{fig:2}
 \end{center}
\end{figure}

On the other hand, the scalar curvature (\ref{ricciscalar}) has the same structure as (\ref{ricciscalar}) up to a $Z_{m}$ factor
\begin{equation}
 R=-\frac{1}{2Z_{m}\left(\lambda-\lambda_{0}\right)} \left[ \sum_{k=0}^{d-2}z_{k}^{2}+\left(\sum_{k=0}^{d-2}z_{k}\right)^{2} \right],\label{Scalarflow}
\end{equation}
and is also negative-definite if $Z_{m}$ is positive-definite, and is zero when $\lambda\mapsto\infty$. In other words, we successfully reach a Ricci-flat space-time in this limit. The fixed points of the flow (\ref{fixedpointdef}) are obtained from the metric (\ref{metric2}) as
\begin{equation}
\begin{array}{rcl}
\partial_{\lambda}g_{rr} &=& 0,\\ \\

\partial_{\lambda}g_{ij} &=& - 2r\eta_{ij}\dfrac{ Z_{m} z_{i}}{\left[ 2Z_{m}\left(\lambda-\lambda_{0}\right)\right]^{3/2}}e^{2\chi z_{i}r}=0,\\ 
\end{array}\label{fixedp}
\end{equation}
which verify the fixed point of space-time when $\lambda\mapsto\infty$. The figure (\ref{fig:2}) shows that under the flow, the evolution of scalar curvature and the homogenization of the geometry are consistent for different values of $Z_{m}$.
Obtained the functions $f(\lambda)$, $g(\lambda)$ and $\chi(\lambda)$, we can analyze the behaviour of the cosmological constant under the Ricci-DeTurck flow. By substituting the $\chi$ expression in the Eq. (\ref{Cosmo}), we have
\begin{equation}
\Lambda=\dfrac{1}{8Z_{m}\left(\lambda-\lambda_{0}\right)} \left\lbrace \dfrac{d-2}{2}\sum_{k=0}z_{k}z_{0}\left[A_{0}^{2}+\left(4-A_{0}^{2}\right)h\right]+\sum_{k=0}\eta^{kk}A_{k}^{2}z_{k}^{2}\right\rbrace,
\end{equation}
which can be taken to be negative during the flow by a suitable relation between vector coefficients and critical exponents. Finally, the masses of the vector fields (\ref{masas}) adopt the form
\begin{equation}
m_{i}^{2}=\dfrac{1}{2Z_{m}\left(\lambda-\lambda_{0}\right)}\left[ \sum_{k=0}z_{k}\left(1-h\right)z_{i}-2z_{i}^{2}\right], \ \ \ \ \forall \ i=0,...,d-2.
\end{equation}
In the limit $\lambda\mapsto\infty$, which corresponds to $\chi\mapsto 0$, these functions converge to zero while the Proca fields tend to a constant, yielding a vanishing entropy functional in concordance with the flat space-time fixed point. Therefore, the matter content of the functional (\ref{action}), the vacuum energy density as well as the scalar curvature become trivial in this limit. Another important result is that the functional (\ref{action}) maintains positive-definite the entropy functional along the whole the Ricci-DeTurck flow.

\section{Discussion and concluding remarks}\label{conclusiones}

In this work, we consider an entropy functional and analyze the behaviour of a family of completely anisotropic Lifshitz metrics under the Ricci-DeTurck flow. We studied the Ricci-DeTurck flow formalism in two scenarios: Einstein-Hilbert and Einstein-Proca with multiple vector fields and negative cosmological constant. 
For both setups we encountered a single fixed point corresponding to flat geometry as the flow parameter tends to infinity. 

In the case when multiple massive vector fields are considered, the $\lambda$ parameter of the Ricci-DeTurck flow allows us to analyze the conditions under which the parameters of the vector fields generate a family of geometries with completely arbitrary metric coefficients as $\lambda\in\left(\lambda_{0},\infty\right)$ evolve. 
The analysis considered here has been performed using the so-called DeTurck \emph{trick}, in which the structure of Ricci flow equations (\ref{ricci1}) can be transformed into a strictly parabolic equation (\ref{ricci2}), and allows us to elucidate an unique functional dependency of the DeTurck vector $V_{r}$. 
By studying the flat space-time fixed point we prove that the entropy functional defined by the Einstein-Proca theory with a negative cosmological constant vanishes as $\lambda$ grows since the scalar curvature, the cosmological constant as well as the mass terms of the Proca fields vanish, while the Proca fields adopt constant values at this point, rendering a trivial Faraday tensor for each field. 
Interestingly enough, for the particular case in which the vector fields are trivial, treated in section (\ref{riccinomattermain}), we still have a Ricci-DeTurck flow of strictly geometric type with the same fixed point. 
Besides, along the flow the scalar curvature displays an increasing monotonous behavior: it starts from infinite negative values at certain value of the flow parameter, increases as the flow goes by, and asymptotically vanishes, in concordance with the expected geometric homogenization picture.

From a theoretical perspective, the entropy functional (\ref{entropy}) can decrease with the flow for general pseudo-Riemannian metrics. The term entropy suggests that we can use it to explore black hole Thermodynamics considering a modification of Perelman's geometric entropy to obtain Ricci solitons, as in \cite{Samuel:2007zz}. Ricci solitons are gradients of suitable scalar fields which modify the Ricci flow and closely resemble the DeTurck vector. With the aid of deformations due to matter, it is interesting to explore metric configurations with black holes and their respective thermodynamic entropy. The behavior of entropy as a function of radial coordinate is central in bulk geometries with holographic dual like Reissner-Nordstr\"om-AdS and Lifshitz black holes. The exact relation between the holographic entropy and the entropy functional is one of the subjects of investigation that we want to explore.

In turn, in the same direction, the family of anisotropic metrics is an important geometric framework in the context of non-relativistic holographic duality in which the scaling symmetry is different for each coordinate. This type of anisotropic scaling symmetry is believed to be present in dual quantum critical points of Condensed Matter Theory, being the Lifshitz space-time the most studied case. In these theories, the critical exponents codify the second-order transitions for physical condensed matter quantities like the density of Cooper electrons in superconducting states.

In this work, we were able to write the full anisotropic metric as a function of these critical exponents, which correspond to the respective scalings of each coordinate of the holographic dual field theory, together with a flowing structure with respect to the $\lambda$ parameter, getting closer to a general description of non-relativistic holographic systems with Ricci flow. In this respect, it has been reported that the Ricci flow is the Holographic Renormalization Group flow \cite{deBoer:1999tgo}. The latter needs to define the stress tensor for the gravity bulk space-time defined in an $r$ constant surface, using the $(d + 1)$ decomposition. The Hamilton equation for the canonical momenta is the Ricci flow but with the radial coordinate as a flow parameter. These canonical momenta are holographically dual to the induced energy-momentum tensor of the field theory defined on that surface. An interesting question is whether the DeTurck vector has meaning in this holographic construction knowing that in $(d + 1)$ decomposition, corresponds to a particular choice of phase vector (with the corresponding lapse vector). The freedom of the choice of phase vector must be reflected on the diffeomorphism invariance of the DeTurck vector. As far as the authors know, this question has not been considered yet and is a motivation for future work.

%\section*{Declaration of competing interest}
%The authors declare that they have no known competing financial interests or personal relationships that could have appeared to influence the work reported in this paper.

\section*{Acknowledgments}
All authors have been benefited from the CONACYT grant No. A1-S-38041, while RCF and AHA were supported by the VIEP-BUAP grant No. 122. 
MCL thanks the financial assistance provided by a CONACYT postdoctoral grant No. 30563.
JAHM acknowledges support from CONACYT through a PhD grant No. 750974. DFHB is also grateful to CONACYT for a Postdoc por M\'exico grant No. 372516.

\newpage
\bibliography{Protocolo}

\providecommand{\href}[2]{#2}\begingroup\raggedright\begin{thebibliography}{10}

\bibitem{10.4310/jdg/1214436922}
R.~S. Hamilton, ``{Three-manifolds with positive Ricci curvature},'' {\em
  Journal of Differential Geometry} {\bf 17} (1982), no.~2, 255 -- 306.

\bibitem{Headrick:2006ti}
M.~Headrick and T.~Wiseman, ``{Ricci flow and black holes},'' {\em Class.
  Quant. Grav.} {\bf 23} (2006) 6683--6708,
  \href{http://www.arXiv.org/abs/hep-th/0606086}{{\tt hep-th/0606086}}.

\bibitem{Samuel:2007zz}
J.~Samuel and S.~R. Chowdhury, ``{Geometric flows and black hole entropy},''
  {\em Class. Quant. Grav.} {\bf 24} (2007) F47--F54,
  \href{http://www.arXiv.org/abs/0711.0428}{{\tt 0711.0428}}.

\bibitem{Samuel:2007ak}
J.~Samuel and S.~R. Chowdhury, ``{Energy, entropy and the Ricci flow},'' {\em
  Class. Quant. Grav.} {\bf 25} (2008) 035012,
  \href{http://www.arXiv.org/abs/0711.0430}{{\tt 0711.0430}}.

\bibitem{Husain:2008rg}
V.~Husain and S.~S. Seahra, ``{Ricci flows, wormholes and critical
  phenomena},'' {\em Class. Quant. Grav.} {\bf 25} (2008) 222002,
  \href{http://www.arXiv.org/abs/0808.0880}{{\tt 0808.0880}}.

\bibitem{Cartas-Fuentevilla:2017cvt}
R.~Cartas-Fuentevilla, A.~Herrera-Aguilar, and J.~A. Olvera-Santamar\'ia,
  ``{Evolution and metric signature change of maximally symmetric spaces under
  the Ricci flow},'' {\em Eur. Phys. J. Plus} {\bf 133} (2018), no.~6, 235,
  \href{http://www.arXiv.org/abs/1707.07235}{{\tt 1707.07235}}.

\bibitem{Cartas-Fuentevilla:2018rez}
R.~Cartas-Fuentevilla, A.~Herrera-Aguilar, and J.~A. Herrera-Mendoza,
  ``{Constructing Lifshitz spaces using the Ricci flow},'' {\em Annals Phys.}
  {\bf 415} (2020) 168093, \href{http://www.arXiv.org/abs/1812.06239}{{\tt
  1812.06239}}.

\bibitem{perelman}
G.~{Perelman}, ``{The entropy formula for the Ricci flow and its geometric
  applications},'' {\em arXiv Mathematics e-prints} (Nov., 2002) math/0211159,
  \href{http://www.arXiv.org/abs/math/0211159}{{\tt math/0211159}}.

\bibitem{10.4310/jdg/1214509286}
D.~M. DeTurck, ``{Deforming metrics in the direction of their Ricci tensors},''
  {\em Journal of Differential Geometry} {\bf 18} (1983), no.~1, 157 -- 162.

\bibitem{deBoer:1999tgo}
J.~de~Boer, E.~P. Verlinde, and H.~L. Verlinde, ``{On the holographic
  renormalization group},'' {\em JHEP} {\bf 08} (2000) 003,
  \href{http://www.arXiv.org/abs/hep-th/9912012}{{\tt hep-th/9912012}}.

\bibitem{PhysRevLett.45.1057}
D.~Friedan, ``Nonlinear models in $2+\ensuremath{\epsilon}$ dimensions,'' {\em
  Phys. Rev. Lett.} {\bf 45} (Sep, 1980) 1057--1060.

\bibitem{Heemskerk:2010hk}
I.~Heemskerk and J.~Polchinski, ``{Holographic and Wilsonian Renormalization
  Groups},'' {\em JHEP} {\bf 06} (2011) 031,
  \href{http://www.arXiv.org/abs/1010.1264}{{\tt 1010.1264}}.

\bibitem{Akhmedov:1998vf}
E.~T. Akhmedov, ``{A Remark on the AdS / CFT correspondence and the
  renormalization group flow},'' {\em Phys. Lett. B} {\bf 442} (1998) 152--158,
  \href{http://www.arXiv.org/abs/hep-th/9806217}{{\tt hep-th/9806217}}.

\bibitem{Balasubramanian:1999jd}
V.~Balasubramanian and P.~Kraus, ``{Space-time and the holographic
  renormalization group},'' {\em Phys. Rev. Lett.} {\bf 83} (1999) 3605--3608,
  \href{http://www.arXiv.org/abs/hep-th/9903190}{{\tt hep-th/9903190}}.

\bibitem{Faulkner:2010jy}
T.~Faulkner, H.~Liu, and M.~Rangamani, ``{Integrating out geometry: Holographic
  Wilsonian RG and the membrane paradigm},'' {\em JHEP} {\bf 08} (2011) 051,
  \href{http://www.arXiv.org/abs/1010.4036}{{\tt 1010.4036}}.

\bibitem{Mikula:2019tyj}
P.~Mikula, M.~E. Carrington, and G.~Kunstatter, ``{Nonequilibrium approach to
  holographic superconductors using gradient flow},'' {\em Phys. Rev. D} {\bf
  100} (2019), no.~4, 046004, \href{http://www.arXiv.org/abs/1902.08669}{{\tt
  1902.08669}}.

\bibitem{Hartnoll:2016apf}
S.~A. Hartnoll, A.~Lucas, and S.~Sachdev, ``{Holographic quantum matter},''
  \href{http://www.arXiv.org/abs/1612.07324}{{\tt 1612.07324}}.

\bibitem{Kachru:2008yh}
S.~Kachru, X.~Liu, and M.~Mulligan, ``{Gravity duals of Lifshitz-like fixed
  points},'' {\em Phys. Rev. D} {\bf 78} (2008) 106005,
  \href{http://www.arXiv.org/abs/0808.1725}{{\tt 0808.1725}}.

\bibitem{Taylor:2008tg}
M.~Taylor, ``{Non-relativistic holography},''
  \href{http://www.arXiv.org/abs/0812.0530}{{\tt 0812.0530}}.

\bibitem{Cartas-Fuentevilla:2018sbl}
R.~Cartas-Fuentevilla, A.~Herrera-Aguilar, V.~Matlalcuatzi-Zamora, U.~Noriega,
  and J.~M. Romero, ``{Anisotropic Lifshitz holography in Einstein-Proca theory
  with stable negative mass spectrum},'' {\em Eur. Phys. J. Plus} {\bf 135}
  (2020), no.~2, 155, \href{http://www.arXiv.org/abs/1804.02278}{{\tt
  1804.02278}}.

\bibitem{Gregory:2010gx}
R.~Gregory, S.~L. Parameswaran, G.~Tasinato, and I.~Zavala, ``{Lifshitz
  solutions in supergravity and string theory},'' {\em JHEP} {\bf 12} (2010)
  047, \href{http://www.arXiv.org/abs/1009.3445}{{\tt 1009.3445}}.

\bibitem{Braviner:2011kz}
H.~Braviner, R.~Gregory, and S.~F. Ross, ``{Flows involving Lifshitz
  solutions},'' {\em Class. Quant. Grav.} {\bf 28} (2011) 225028,
  \href{http://www.arXiv.org/abs/1108.3067}{{\tt 1108.3067}}.

\bibitem{Woolgar:2007vz}
E.~Woolgar, ``{Some Applications of Ricci Flow in Physics},'' {\em Can. J.
  Phys.} {\bf 86} (2008) 645--651,
  \href{http://www.arXiv.org/abs/0708.2144}{{\tt 0708.2144}}.

\end{thebibliography}\endgroup

\bibliographystyle{./utphys}

\end{document}